\documentclass[11pt,fleqn]{article}
\usepackage{amssymb, euscript, amsmath}
\usepackage{color}

\newtheorem{thm}{Theorem}

\begin{document}


\title{\protect\vspace*{-1cm}\bf The four-dimensional Mart{\'{\i}}nez Alonso--Shabat equation:\\ reductions  and nonlocal symmetries}

\author{Oleg I. Morozov$^1$\footnote{Present address: Faculty of Applied Mathematics, AGH University of Science and Technology, Al. Mickiewicza 30, 30059 Krak\'ow, Poland. E-mail: {\tt morozov@agh.edu.pl}}\ \ and Artur Sergyeyev$^2$\\
%
%
$^1$Institute of Mathematics and Statistics, University of Troms\o, \\ Troms\o \, 90-37, Norway\\
$^2$Mathematical Institute, Silesian University in Opava, \\Na Rybn\'\i{}\v{c}ku 1, 746\,01 Opava, Czech Republic\\
E-mail: {\tt oleg.morozov@uit.no} and {\tt artur.sergyeyev@math.slu.cz}}

\date{\protect\vspace*{-0.7cm}}
\maketitle

\begin{abstract}\vspace*{-1cm}We consider the four-dimensional integrable Mart{\'{\i}}nez Alonso--Shabat equation, and list three integrable
three-dimensional reductions thereof. We also present a four-dimensional integrable modified Mart{\'{\i}}nez Alonso--Shabat equation together with its Lax pair.
\looseness=-1

We also construct an infinite hierarchy of commuting nonlocal symmetries (and not just the shadows, as it is usually
the case in the literature) for the Mart{\'{\i}}nez Alonso--Shabat equation.\looseness=-1

\smallskip


\smallskip

\protect\vspace*{-0.3cm}
\end{abstract}






\large
\section{Introduction}

Consider the four-dimensional Mart{\'{\i}}nez Alonso--Shabat equation\looseness=-1
\begin{equation}
u_{ty} = u_z\,u_{xy}-u_y\,u_{xz}
\label{MASh}
\end{equation}
introduced in \cite{as}. It has \cite{m0} a covering defined by system
\begin{equation}
q_y=\lambda\, u_y\, q_x,
\qquad
q_z=\lambda\, (u_z \,q_x - q_t)
\label{lax}
\end{equation}
with a non-removable parameter $\lambda \neq 0$, and a recursion operator, and is therefore integrable.

Below we present three reductions of (\ref{MASh}) to integrable three-dimensional equations: the so-called rdDym
equation \cite{blaszak,pavlov,m1,Ovsienko2010}, the universal hierarchy equation \cite{as}, and an equation
(\ref{eqr}) related \cite{AdlerShabat,fm} to the ABC equation, see \cite{zakharevich,d} for the
latter.

Note that eliminating $u$ from the Lax pair (\ref{lax})
for (\ref{MASh}) yields an integrable four-dimensional PDE (\ref{4D_Veronese_web_eq})
to which we refer to as to the {\em modified} Mart{\'{\i}}nez Alonso--Shabat equation and which can be seen as
a four-dimensional generalization of the ABC equation (\ref{qeqr}).
Integrability of (\ref{4D_Veronese_web_eq})
is established by presenting a Lax pair (\ref{Zakharevich_BT_4D}) for the latter.
It would be interesting to study (\ref{4D_Veronese_web_eq}) in more detail, e.g.\ to find
a recursion operator for this equation.

The main goal of the rest of the present paper is
to find nonlocal symmetries for equation (\ref{MASh}) in the covering (\ref{S}) derived from
the Lax pair (\ref{lax}).

Following \cite{vin4, vkl, olver_eng2} and references therein, recall that a {\em higher} (or {\em generalized} \cite{olver_eng2}) {\em symmetry} for a partial differential
system $\EuScript{E}$ can be identified with its characteristics, which is, roughly speaking,
a vector function on an appropriate jet space $J^\infty(\EuScript{E})$
associated with $\EuScript{E}$ that depends on finitely many arguments
and satisfies the linearized version of $\EuScript{E}$.

Next, a (differential) {\it covering} for a partial differential
system $\EuScript{E}$ is (see e.g.\ \cite{vin0,vkl,vin2, kv})
an over-determined system $\tilde{\EuScript{E}}$ involving additional dependent variables (which are
called {\it pseudopotentials}) such that $\EuScript{E}$ implies the compatibility conditions of
$\tilde{\EuScript{E}}$. A vector function on $J^\infty(\tilde{\EuScript{E}})$ that depends on finitely many arguments
and satisfies the linearized version of $\EuScript{E}$ is called \cite{vin0,vin1,vin2, kv} a {\em shadow} (more precisely, a $\tilde{\EuScript{E}}$-shadow) for $\EuScript{E}$.

Thus, in contrast with the symmetry of $\EuScript{E}$, a shadow is
allowed to depend on the pseudopotentials and their  derivatives.
%
On the other hand, symmetries of $\tilde{\EuScript{E}}$ are called \cite{vin0,vin1,vin2,vkl,kv}
{\it nonlocal symmetries}
of $\EuScript{E}$ associated with the covering $\tilde{\EuScript{E}}$.

Infinite-dimensional Lie algebras of nonlocal symmetries are well known to play
an important role in the theory of integrable systems and provide a useful
tool for the study of the latter, see e.g.\ \cite{blaszak, kv} and references therein.
In this connection 
note (see e.g.\ \cite{vin2}) that not every $\tilde{\EuScript{E}}$-shadow can be lifted to
a full-fledged nonlocal symmetry for $\EuScript{E}$ associated with $\tilde{\EuScript{E}}$.
This fact has profound consequences.
In particular, while for the full nonlocal symmetries
we can readily define their Lie bracket, this is not quite the case for the shadows.
\looseness=-1

The results of Sections~\ref{cov} and \ref{nls} of the present paper can now be summarized as follows.

First, using (\ref{lax}) as a starting point we construct a new covering (\ref{S}) for (\ref{MASh}).
Expanding the pseudopotential in (\ref{S}) into the formal Taylor series w.r.t.\ the spectral parameter $\lambda$
gives a new covering (\ref{Wi}) with an infinite number of new pseudopotentials $w_i$ which are the coefficients at the powers of
$\lambda$.\looseness=-1

We then proceed to construct an infinite hierarchy of commuting nonlocal symmetries for equation (\ref{MASh}) in this new
covering using a technique from \cite{s09}. Let us stress that this construction makes heavy use of the fact that the covering (\ref{Wi})
can be promoted to a covering (\ref{Wi})+(\ref{taui2-w}) over the system
that consists of (\ref{MASh}) and (\ref{taui2-u}).
It should also be mentioned that finding full-fledged nonlocal symmetries for
equations with more than two independent variables rather than mere shadows
is quite uncommon, cf.\ e.g.\ \cite{fmi, s09} and the discussion at the end of Section~\ref{nls}.

\section{Reductions of the Mart{\'{\i}}nez Alonso--Shabat equation}\label{red}


It is a remarkable fact that three known integrable three-dimensional PDEs can be obtained as reductions of (\ref{MASh}).

\subsection{The rdDym equation}

The reduction $z=x$ yields the rdDym equation \cite{blaszak,pavlov,m1,Ovsienko2010} that arises as the $r\rightarrow\infty$ limit of the so-called $r$th dispersionless Harry Dym equation \cite{blaszak}:\looseness=-1
\begin{equation}
u_{ty} = u_x \,u_{xy} - u_y \,u_{xx}.
\label{rdDym}
\end{equation}
The Lax pair (\ref{lax}) after the reduction boils down to the known Lax pair for (\ref{rdDym}),
\[
q_t = (u_x - \lambda^{-1}) \, q_x,
\qquad
q_y = \lambda \, u_y\,q_x.
\]

\subsection{The universal hierarchy  equation}

Putting $t=y$ in (\ref{MASh}) and (\ref{lax}) yields the universal hierarchy equation \cite{as}
\[
u_{yy} = u_z \, u_{xy} - u_y \, u_{xz}
\]
and its Lax pair
\[
q_y = \lambda \, u_y \, q_x,
\qquad
q_z = \lambda \, (u_z - \lambda \, u_y) \, q_x.
\]

\subsection{An equation related to the ABC equation}

Another interesting reduction admitted by (\ref{MASh}) arises when we put $z=t$. This produces the equation
\begin{equation}
u_{ty}= u_t\, u_{xy} - u_y\, u_{tx},
\label{eqr}
\end{equation}
which, along with the associated
Lax representation,
\begin{equation}
q_t=\lambda\, u_t\, q_x/(\lambda+1),
\qquad
q_y=\lambda \,u_y\, q_x,
\label{laxr}
\end{equation}
obtained by performing the reduction in question in (\ref{lax}), has already appeared in the literature,
see \cite{AdlerShabat,fm}.\looseness=-1

Upon eliminating $u$ from (\ref{laxr}) we arrive at the equation
\begin{equation}
q_y \,q_{tx}=(\lambda+1)\, q_t\, q_{xy} -\lambda \,q_x \, q_{ty},
\label{qeqr}
\end{equation}
which is, up to the removal of non-essential parameters,
nothing but the so-called ABC equation
\begin{equation}\label{abc}
A\, q_x \, q_{ty} +B\, q_y\, q_{tx} + C \, q_t \, q_{xy}=0,
\qquad A+B+C =0,
\end{equation}
which describes three-dimensional Veronese webs \cite{zakharevich, d}.
This equation is also of importance in geometry: as shown in \cite{fk}, to any smooth solution of (\ref{abc}) one can associate a three-dimensional Einstein--Weyl structure. The B\"acklund transformation
(\ref{laxr}) relating (\ref{eqr}) and (\ref{qeqr}) appears in Remark~2 of \cite{fm}.
%
In \cite{bft} it was shown that (\ref{abc}) with $A+B+C\neq 0$ is
also integrable but has a nonisospectral Lax pair.

As a final remark, note that the equations (\ref{qeqr}) and
\begin{equation}
r_y \,r_{tx}=(\mu+1)\, r_t\, r_{xy} -\mu \, r_x \, r_{ty},
\label{r_qeqr}
\end{equation}
with the parameters $\mu \neq \lambda$ are related by the following B\"acklund transformation \cite{zakharevich}:
\begin{equation}
r_t = \frac{\mu\,(\lambda+1)}{\lambda\,(\mu+1)}\,\frac{q_t}{q_x}\,r_x,
\qquad
r_y = \frac{\mu}{\lambda}\,\frac{q_y}{q_x}\,r_x.
\label{Zakharevich_BT}
\end{equation}


\section{The modified Mart{\'{\i}}nez Alonso--Shabat equation}

Eliminating $u$ from (\ref{lax}) yields an equation
\begin{equation}
q_y\,q_{xz} + \lambda \, q_x \, q_{ty} - (q_z+\lambda\, q_t)\, q_{xy} = 0,
\label{4D_Veronese_web_eq}
\end{equation}
which involves a parameter $\lambda$ and can be considered as a four-dimensional
generalization of the ABC equation (\ref{qeqr}) and reduces to the latter if $z=t$.

Moreover, there exists
a 4D generalization of (\ref{Zakharevich_BT}): for $\mu \neq \lambda$ the system
\begin{equation}
r_y = \frac{\mu}{\lambda}\,\frac{q_y}{q_x}\,r_x,
\qquad
r_z = \frac{\mu}{\lambda}\,\frac{q_z+\lambda \, q_t}{q_x}\,r_x - \mu \, r_t
\label{Zakharevich_BT_4D}
\end{equation}
defines a B\"acklund transformation between equations (\ref{4D_Veronese_web_eq}) and
\[
r_y\,r_{xz} + \mu \, r_x \, r_{ty} - (r_z+\mu\,  r_t)\,r_{xy} = 0.
\]
This {\em inter alia} means that (\ref{Zakharevich_BT_4D}) is compatible by virtue of (\ref{4D_Veronese_web_eq}), and thus (\ref{Zakharevich_BT_4D}) provides a Lax pair for (\ref{4D_Veronese_web_eq}) with the spectral parameter $\mu$.
Thus, (\ref{4D_Veronese_web_eq}) is a four-dimensional integrable equation, to which we refer as to the {\em modified Mart{\'{\i}}nez Alonso--Shabat equation}.

\section{New coverings for the Mart{\'{\i}}nez Alonso--Shabat equation}
\label{cov}

Consider the Lax operators associated with the Lax pair (\ref{lax}),
\[
L_1=D_z+\lambda\,(D_t-u_z\, D_x),\qquad L_2=D_y-\lambda \,u_y \,D_x,
\]
where $D_t$, $D_x$, $D_y$, $D_z$  are total derivatives in the covering (\ref{lax}) over (\ref{MASh}).
We can now construct another covering over (\ref{MASh}) as follows. Let $M=s \, D_x$.
Then it is readily checked that the equations $[L_i,M]=0$, $i=1,2$,
boil down to the following equations for $s$: \looseness=-1
\begin{equation}
s_y=\lambda \, (u_y \,s_x- u_{xy}\, s),
\qquad
s_z=\lambda(u_z\, s_x-s_t -u_{xz}\, s).
\label{S}
\end{equation}
These equations are compatible by virtue of (\ref{MASh}),
and thus
define a covering over (\ref{MASh}) with the pseudopotential $s$.


The covering (\ref{S}) is of interest {\em inter alia} for the following reason: it is easily checked that $U=s$
satisfies the linearized version of (\ref{MASh})
\begin{equation}
\label{lineq}
U_{ty}=u_z \,U_{xy} -u_y\, U_{xz}+u_{xy}\, U_z- u_{xz}\,U_y,
\end{equation}
i.e., $s$ is a shadow for (\ref{MASh}) in this covering.

To any shadow one can associate a covering in which this shadow is lifted to a nonlocal symmetry using the construction of \cite{Khorkova, Kiso}. However, this requires an introduction of an additional infinite series of nonlocal variables, and these series are different for different shadows.

It is therefore a remarkable fact that for equation (\ref{MASh}) we were able to construct a fairly simple covering (\ref{Wi})
for which there exists an infinite commuting series of nonlocal symmetries of (\ref{MASh}) expressible solely in terms of pseudopotentials $w_i$ of this covering. The technique employed by us to this end below mimics the one from \cite{s09}.

Namely, following \cite{s09}, consider a copy of the covering (\ref{S}) with $\lambda$ replaced by another parameter $\mu$ and the pseudopotential denoted by $w$ instead of $s$:
\begin{equation}
w_y=\mu \,(u_y\, w_x - u_{xy}\, w),
\qquad
w_z=\mu \, (u_z\, w_x - w_t -u_{xz} w).
\label{W}
\end{equation}
By the above, $w$ is a shadow of nonlocal symmetry for (\ref{MASh}).
Informally this can be restated as follows: suppose that $u$, $s$, $w$
also depend on an additional independent variable $\tau$, then equation $u_{\tau} = w$
is compatible with (\ref{MASh}), (\ref{W}).\looseness=-1

Moreover, it turns out that the 
system
\begin{equation}
u_{\tau}=w,
\qquad
s_\tau=\displaystyle\frac{\lambda\,\mu}{\mu-\lambda}\,(w \, s_x - s \, w_x)
\label{tau}
\end{equation}
is compatible with (\ref{MASh}), (\ref{S}) by virtue of (\ref{MASh}), (\ref{S}), (\ref{W}) and (\ref{tau}).
It is important to stress that the extension to $s$ of the first part of the flow, $u_{\tau}=w$, is not uniquely defined, see e.g.\ \cite{vin2}, i.e., the choice made for the right-hand side of the second equation of (\ref{tau}) is, generally speaking, not a canonical one.\looseness=-1

Now slightly alter the notation to stress the dependence of all relevant objects on $\mu$:
write $\tau(\mu)$ instead of $\tau$ and $w(\mu)$ instead of $w$. In this notation $s=w(\lambda)$, so if we assume, following (\ref{tau}), that there holds
\begin{equation}\label{formax0}
(w(\mu))_{\tau(\nu)}=\displaystyle\frac{\mu\,\nu}{\nu-\mu}\,(w(\nu) \, (w(\mu))_x - w(\mu) \, (w(\nu))_x),
\end{equation}
and its counterpart with $\mu$ and $\nu$ interchanged,
then the flows with the times $\tau(\mu)$ and $\tau(\nu)$ are readily checked to commute for all $\mu\neq \nu$:
\begin{equation}
\frac{\partial^2 u}{\partial \tau(\mu)\partial \tau(\nu)}=\frac{\partial^2 u}{\partial \tau(\nu)\partial \tau(\mu)},
\qquad
\frac{\partial^2 s}{\partial \tau(\mu)\partial \tau(\nu)} =\frac{\partial^2 s}{\partial \tau(\nu)\partial \tau(\mu)}.
\label{com}
\end{equation}

As an aside note that if we put $\tilde{w}(\mu)=1/(\mu \,w(\mu))$, then (\ref{formax0}) can be written in the conservative form as
\begin{equation}\label{formax1}
(\tilde{w}(\mu))_{\tau(\nu)}=\displaystyle\frac{\mu}{\nu-\mu}\,(\tilde{w}(\mu)/\tilde{w}(\nu))_x.
\end{equation}
Up to passing to the inverses of the parameters ($\mu'=1/\mu$, $\nu'=1/\nu$)
and changing the sign of $\tau(\nu)$ and some bits of notation,
this is 
nothing but equation (20) from \cite{pavlov}, i.e., the equation for the generating
function 
for the conserved densities of the so-called $\varepsilon$-system,
see \cite{pavlov} for further details and cf.\ also \cite{AdlerShabat,Shabat2012}.\looseness=-1

Expanding $w(\mu)$ into formal power series in $\mu$, $w(\mu)=\sum\limits_{i=0}^{\infty}w_i\mu^i$, now yields
a new covering over (\ref{MASh}) with the pseudopotentials $w_i$ defined by the system
\[
(w_{i})_{y}=u_y\, (w_{i-1})_x - u_{xy}\, w_{i-1},
\]
\begin{equation}
(w_{i})_{z}=u_z\, (w_{i-1})_x - (w_{i-1})_t -u_{xz} \, w_{i-1}
\label{Wi}
\end{equation}
for $i \in \mathbb{N}$ with an arbitrary smooth function $w_0=w_0(t,x)$.

The pseudopotentials $w_i$, $i\in\mathbb{N}$, are readily checked to be shadows of nonlocal symmetries of equation (\ref{MASh}) in the covering (\ref{Wi}).  It turns out that we reproduce the action of the inverse recursion operator $\mathcal{R}^{-1}$ for (\ref{MASh}) from \cite{m0}: we have $w_i=\mathcal{R}^{-1} w_{i-1}$, $i\in\mathbb{N}$.\looseness=-1

\section{Nonlocal symmetries of the Mart{\'{\i}}nez Alonso--Shabat equation}\label{nls}
We now intend to show that the shadows $w_i$ can be lifted to full-fledged nonlocal symmetries for (\ref{MASh}) in the sense of \cite{vin0,vin1,vin2}. To this end we write, following the spirit of theory of generating functions for commuting flows, cf.\ e.g.\  equation (4) in \cite{kp}, and references therein, a formal expansion
\[
\frac{\partial}{\partial\tau(\mu)}=\sum\limits_{i=0}^\infty \mu^i \, \frac{\partial}{\partial\tau_i},
\]
and substitute this, along with the formal expansion for $w$ in $\mu$, into (\ref{tau}).

This results in the following equations:
\begin{equation}
u_{\tau_i}=w_i,
\qquad
s_{\tau_i}=\displaystyle\sum\limits_{k=0}^{i-1} \lambda^{k-i+1}((w_{k})_{x}\, s  - w_k \, s_x),\quad i \in \mathbb{N}.
\label{taui}
\end{equation}
It is readily checked that for any given $i\in\mathbb{N}$ the system (\ref{taui}) is
compatible with (\ref{MASh}), (\ref{S}) by virtue of (\ref{MASh}), (\ref{S}), (\ref{Wi}) and (\ref{taui}).

As $s\equiv w(\lambda)$, we also have a formal expansion 
$s=\sum\limits_{j=0}^{\infty}w_j\lambda^j$.
Substituting this into the second equation of (\ref{taui}) yields the system
\begin{subequations}
\label{taui2}
\begin{align}
u_{\tau_i}&=w_i,\label{taui2-u}\\
(w_{j})_{\tau_i}&=\sum\limits_{k=0}^{i-1} (w_{i+j-k-1} \,(w_{k})_{x} - w_k \, (w_{i+j-k-1})_{x}),
\label{taui2-w}
\end{align}
\end{subequations}
which is compatible with (\ref{MASh}) and (\ref{Wi}); here $i, j \in \mathbb{N}$.

It is easily seen that the flows (\ref{taui2}) commute, i.e., for all $i, j, k \in \mathbb{N}$ we have
\begin{equation}
\frac{\partial^2 u}{\partial \tau_i\partial \tau_j}=\frac{\partial^2 u}{\partial \tau_j\partial \tau_i},
\qquad
\frac{\partial^2 w_k}{\partial \tau_i\partial \tau_j}=\frac{\partial^2 w_k}{\partial \tau_j\partial \tau_i}.
\label{comtau}
\end{equation}
In fact, this result can be extracted directly from (\ref{com}).

Thus, we arrive at the following theorem, which can also be proved by direct computation
instead of the above reasoning:

\begin{thm}\label{theocom}
The infinite prolongations of the vector fields
\begin{equation}
Q_i = w_i \,\displaystyle\frac{\partial}{\partial u}
+\displaystyle \sum\limits_{j=1}^{\infty} \sum\limits_{k=0}^{i-1}
 (w_{i+j-k-1} \,(w_{k})_{x} - w_k \, (w_{i+j-k-1})_{x})
\,\frac{\partial}{\partial w_j},
\label{Qi}
\end{equation}
where $i \in \mathbb{N}$, upon restriction to
(\ref{MASh}) and (\ref{Wi}) form an infinite series of commuting nonlocal symmetries for equation (\ref{MASh}) in the covering (\ref{Wi}).
\end{thm}

The commutativity of $Q_i$ in this context means that the Lie brackets of the infinite prolongations of $Q_i$
(or, equivalently, the Jacobi brackets of their characteristics, cf.\ \cite{vin2}) vanish for all $i$.
This essentially amounts to (\ref{comtau}).

Note that existence of an infinite hierarchy of commuting flows
is one of the most important hallmarks of integrability, see e.g.\
\cite{ff, act, vkl, olver_eng2, blaszak-book, ms} and references therein.\looseness=-1

As we have $w_i=\mathcal{R}^{-1} w_{i-1}$, $i\in\mathbb{N}$, where $\mathcal{R}$ is the recursion operator for (\ref{MASh}) found in \cite{m0}, cf.\ above, the commutativity (\ref{comtau}) of the flows (\ref{taui}) suggests that $\mathcal{R}^{-1}$ and $\mathcal{R}$ should be hereditary (cf.\ e.g.\ \cite{ff, blaszak-book, olver_eng2} for more details on this property) in some appropriate sense, at least on the linear span of $w_i$. It is not quite clear, however, whether this claim can be made precise, let alone proved, because of the complicated structure of nonlocal terms in $\mathcal{R}^{-1}$ and $\mathcal{R}$.\looseness=-1

To conclude, let us stress again that finding an explicit form of the generators and the commutation relations for an infinite-dimensional algebra of nonlocal symmetries (rather than just the shadows, cf.\ e.g.\ \cite{vin0,vin1,vkl,vin2} for the details on differences among the two) for a non-overdetermined multidimensional PDE is quite rare. We were able to find just two similar results in the literature: the first is the commutative self-dual Yang--Mills hierarchy \cite{act}, and the second \cite{fmi} is an infinite-dimensional noncommutative algebra of nonlocal symmetries for the (2+1)-dimensional integrable Boyer--Finley equation. It would therefore be very interesting to obtain the results similar to our Theorem~\ref{theocom} for other multidimensional integrable systems, for instance, those recently found in \cite{aserg}.
\looseness=-1

\subsection*{Acknowledgments}

This research of AS was supported in part by the Ministry of
Education, Youth and Sports of the Czech Republic (M\v{S}MT \v{C}R)
under RVO funding for I\v{C}47813059, and by the Grant Agency of the Czech Republic
(GA \v{C}R) under grant P201/12/G028.

This research was initiated in the course of visits
of OIM to the Silesian University in Opava and of AS to the University of Troms{\o}.
Both authors thank the universities they visited for the warm hospitality extended to them.

The authors also thank E.V. Ferapontov, M.V. Pavlov,
and the anonymous referee for useful suggestions.

\end{document}